# On topology optimization of acoustic metamaterial lattices for locally resonant bandgaps of flexural waves

Saeid Hedayatrasa[*], Kazem Abhary and Mohammad Uddin

School of Engineering, University of South Australia, Mawson Lakes, SA 5095, Australia

*Email: saeid.hedayatrasa@mymail.unisa.edu.au

(Dated: November 17, 2016)

**ABSTRACT**

Optimized topology of bi-material acoustic metamaterial lattice plates is studied for maximized locally resonant bandgap of flexural guided waves. Optimized layout of the two relatively stiff and compliant material phases in the design domain is explored, free from any restrictions on the topology and shape of the relevant domains. Multiobjective optimization is performed through which maximized effective stiffness or minimized overall mass of the bandgap topology is additionally ensured. Extreme and selected intermediate optimized topologies of Pareto fronts are presented and their bandgap efficiencies and effective stiffness are compared. The bi-material constitution of selected topologies are further altered and modal band structure of resultant multilateral and porous designs are evaluated. Novel, core-shell like, locally resonant bandgaps are introduced. It is shown that how the bandgap efficiency and structural mass and/or stiffness can be optimized through optimized microstructural design of the matrix and the resonating core domains.

## 1. INTRODUCTION

Acoustic metamaterials are composite materials with designed heterogeneity to manipulate vibroacoustic waves for e.g. wave steering, resonation, guiding, and filtration. Acoustic metamaterial lattices can be designed by which vibroacoustic waves are filtered out within particular frequency ranges called acoustic bandgap (ABG). Such ABG materials exponentially decay the wave's amplitude due to the destructive interface of their periodic microstructure. The width and frequency range of bandgap depends on the contrast of constitutive materials and lattice shape and the topology of its irreducible unit-cell. When the wavelength is comparable to the lattice periodicity (i.e. unit-cell size), then constructive reflection of wave through a stiff scattering material phase (Bragg reflection) opens so-called phononic bandgap. However, it is well-known that relatively low frequencies having wavelength larger than unit-cell size may also be manipulated through localized resonances.

Locally resonant acoustic bandgaps (LRABs) can be produced by introducing periodic resonating features in a background material. These features act as internal oscillators and attenuate wave through destructive out-of-phase oscillations at their local resonance frequency. The oscillations may be dipolar or monopolar leading to negative effective dynamic mass and elastic modulus, respectively. Consequently, a bandgap frequency range is induced over which existence of modal frequencies is banned. Mie resonances of soft-dense scatters in a relatively hard matrix may produce monopole and/or dipole locally resonant bandgaps when the wavelength inside the scatter is comparable to its size (Wang et al., 2004, Li and Chan, 2004, Hsu and Wu, 2007). Core-shell LRABs, with dipole resonance, can be produced by insertion of a stiff-dense core coated with a soft intermediate material in a relatively stiff matrix (Liu et al., 2000). Such LRABs but with single or bi-material constitution may also be designed if the compliance of intermediate coating domain (Bigoni et al., 2013, Wang et al., 2014) and locally resonant features (Wang and Wang, 2013, Yu et al., 2013, Liu et al., 2015) are effectively obtained through appropriate microstructure. LRABs can also be produced by periodic attachment of resonating stubs to a uniform or heterogeneous base plate (Oudich et al., 2010, Oudich et al., 2011, Bilal and Hussein, 2013).

In design of ABGs it is normally desired to achieve broadest bandgap frequency at lowest frequency range through specified unit-cell size. In this way, low frequencies can be manipulated by tiny and possibly subwavelength features. Topology optimization of unit-cell has shown to be a great tool for designing ABGs. However, preceding topology optimization studies have been mostly concerned with design of phononic bandgaps e.g. (Sigmund and Jensen, 2003, Bilal and Hussein, 2011, Hedayatrasa et al., 2016b, Hedayatrasa et al., 2016a). Lu et al. (2013) implemented level set-based topology optimization for optimization of LRABs with unidirectional negative dynamic bulk modulus at given frequency, and further (LU et al., 2014) for negative dynamic mass density. Krushynska et al. (2014) performed a comprehensive parametric study on design of core-shell LRABs and presented how the bandgap efficiency can be simply enhanced by tuning the filling fraction and material properties of various domains; different shapes of the inclusions were also prescribed and compared. Most recently Yang et al. (2016) implemented method of moving asymptotes for topology optimization of the intermediate coating layer of LRABs confined between prescribed matrix and core design domains; first bandgap of in-plane modes was maximized by broadening the relevant frequency range with negative dynamic mass density characteristic.

This paper is motivated to explore optimized topology of bi-material LRABs, free from any restrictions on the topology and shape of the material domains. Optimized layout of the relatively stiff and compliant materials in the design domain is explored for maximized bandgap efficiency while maximizing effective structural stiffness or minimizing overall mass. The compliant material phase can form the soft coating layer and the stiff-dense material phase can form the locally resonant core as well as the hard matrix. The focus is on LRAB of fundamental flexural guided waves in lattice plates which carry a large portion of wave energy, are efficiently coupled with acoustic waves and contribute to sound radiation.

## 2. OPTIMIZATION PROBLEM AND CONSTITUTIVE FORMULATION

Square symmetric plate unit-cell with an aspect ratio (width over thickness) of $a/h = 2$ is modelled to be optimized for LRABs. Bi-material constitution including a soft phase (rubber) and a stiff-dense phase (copper) is assumed, as shown in Figure 1 for an arbitrary topology.

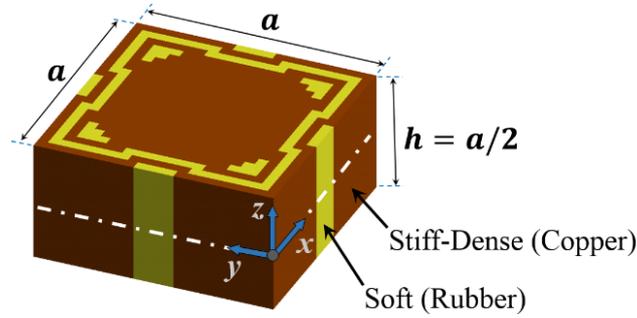

Figure 1 Square symmetric bi-material plate unit-cell with aspect ratio $a/h = 2$ for an arbitrary topology

Wave propagation in such heterogeneous medium is governed by the Navier's equation of equilibrium:

$$\nabla(\lambda + \mu)\nabla \cdot \mathbf{u} + \nabla.\mu\nabla \mathbf{u} = \rho\ddot{\mathbf{u}} \qquad (1)$$

where $\nabla$ is the gradient operator, $\mathbf{u} = \{u \quad v \quad w\}$ is displacement vector, $\rho$ is mass density and $\lambda$ and $\mu$ are the first and second Lame constants respectively. According to the Bloch-Floquet wave theory, the displacement $\mathbf{u}$ in a perfectly periodic plate structure of the unit-cell with 2D lattice periodicity in $xy$-plane can be defined as harmonic modulation of a periodic component:

$$\mathbf{u}(\mathbf{x}, t) = \tilde{\mathbf{u}}(\mathbf{x})e^{(i\mathbf{k}\cdot\mathbf{x}-i\omega t)} \qquad (2)$$

where $\mathbf{x} = \{x \quad y \quad z\}$ is the location vector, $t$ is time, $\omega = 2\pi f$ is circular frequency, $\mathbf{k} = \{k_x \quad k_y\}$ is wave vector corresponding to the 2D lattice periodicity $\mathbf{A} = \{a \quad a\}$, $i = \sqrt{-1}$ and $\tilde{\mathbf{u}}(\mathbf{x}) = \tilde{\mathbf{u}}(\mathbf{x} + \mathbf{A})$ is the periodic component of displacement induced by lattice periodicity. The modal band structure of unit-cell topology is calculated through finite element analysis (FEA) and periodic boundary conditions are applied to the periodic boundaries (parallel to the z-axis) according to the Eq.2 (Hedayatrasa et al., 2016b). Hence for any given wave vector $\mathbf{k}$, the modal frequencies are calculated by eigenvalue analysis of the following equation:

$$|\mathbf{M}^{-1}\mathbf{K}(\mathbf{k}) - \omega^2 \mathbf{I}| = \mathbf{0} \qquad (3)$$

where $\mathbf{M}$ and $\mathbf{K}$ are FEA mass and stiffness matrices, respectively, and $\mathbf{I}$ is the identity matrix. Due to harmonic periodicity of the Eq.2 and square symmetry of unit-cell, modal band structure is calculated by searching discrete points over the edges of irreducible Brillouin zone which is the triangle defined by corners $\Gamma$ ($\mathbf{k} = \{0 \quad 0\}$), X ($\mathbf{k} = \{\pi/a \quad 0\}$) and M ($\mathbf{k} = \{\pi/a \quad \pi/a\}$) in wave vector space. Just half of the unit-cell's thickness ($z = 0$ to $z = h/2$) is modelled and in-plane displacement of the mid-plane is constrained in the FEA model $\{u \quad v\}|_{z=0} = 0$ to decouple asymmetric modal branches (Hedayatrasa et al., 2016b).

The first and primary objective of optimization is defined as maximized relative bandgap width (RBW) of the first two asymmetric modal branches (bandgap width over midgap of eigenvalues):

$$F_1 = \frac{\omega_{2,\min}^2 - \omega_{1,\max}^2}{0.5(\omega_{2,\min}^2 + \omega_{1,\max}^2)} \quad (4)$$

where $\omega_{2,\min}$ is the minimum modal frequency of the second modal branch and $\omega_{1,\max}$ is maximum modal frequency of the first modal branch over the border of irreducible Brillouin zone. Maximizing $F_1$ ensures obtaining widest bandgap at lowest frequency range as desired for optimization of acoustic bandgaps.

The second objective of optimization is defined as maximized in-plane stiffness (i.e. minimized compliance) or minimized mass of the unit-cell and relevant topologies are optimized separately. To achieve maximized stiffness, following earlier work of the authors (Hedayatrasa et al., 2016b) relative compliance of unit-cell is defined to be minimized:

$$F_{2C} = \frac{1}{\epsilon_s}\left(\left(\frac{1-\nu_e}{E_e}\right) + \left(\frac{1}{G_e}\right)\right) \quad (5)$$

where $E_e$, $G_e$ and $\nu_e$ are homogenized orthotropic elastic modulus, shear modulus and Poisson's ratio of the unit-cell defined by standard FEA computational homogenization (Steven, 2006), and constant $\epsilon_s = (1-\nu_s)/E_s + 1/G_s$ is the compliance of pure stiff matrix material (copper).

For the other optimization case, minimized mass of the unit-cell is desired as the second objective and therefore filling fraction of the stiff-dense matrix material is defined to be minimized:

$$F_{2M} = v_f \quad (6)$$

Soft rubber with elastic modulus $E_r = 3.239 \times 10^6$ Pa, Poisson's ratio $\nu_r = 0.4997$ and density $\rho_r = 1050$ kg/m³, and stiff-dense copper with $E_s = 128 \times 10^9$ Pa, $\nu_s = 0.34$ and $\rho_s = 8960$ kg/m³ are used for modelling and optimization of assumed bi-material unit-cell. The high contrast of elastic properties and mass density, and very low elastic wave speed in rubber ($\sqrt{E_r/\rho_r} = 55.54$ m/s) compared with that of copper ($\sqrt{E_s/\rho_s} = 3779.60$ m/s) can lead to various types of acoustic bandgaps. Alternative topologies with potential bandgap characteristic are shown in Figure 2. Bragg scattering of wave at interface of stiff copper scatters in the rubber matrix may open a phononic bandgap (Figure 2(a)). Localized Mie resonances of the soft rubber scatters inside copper matrix can also open a bandgap at relatively low resonance frequency of scatters (Figure 2(b)); this bandgap can be enhanced by addition of an oscillatory copper inclusion inside the rubber scatter (Figure 2(c)) which reduces the local resonance frequency.

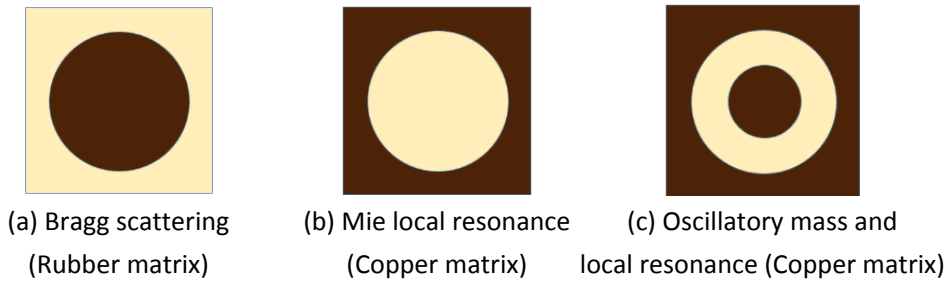

(a) Bragg scattering (Rubber matrix)   (b) Mie local resonance (Copper matrix)   (c) Oscillatory mass and local resonance (Copper matrix)

Figure 2 Schematic presentation of alternative topologies of bi-material unit-cell and dominant bandgap mechanisms

With regards to distinct contrast of the constitutive materials and according to the definition of bandgap objective (Eq.4) which aims at maximizing RBW, the alternative topology mode shown in Figure 2(c) is expected to be promoted during optimization by which lowest bandgap frequency range can be achieved. Following earlier work of the authors (Hedayatrasa et al., 2016b) non-dominated sorting genetic algorithm NSGA-II (Pratap et al., 2002) is adopted for present topology optimization study which provides a Pareto front of optimized solutions with non-dominated magnitudes of the two objective functions.

## 3. RESULTS AND DISCUSSION

Topology resolution of $24 \times 24$ is considered and topology optimization is performed by a uniform in-plane mesh resolution of $48 \times 48$ linear solid cubic elements SOLID185 using ANSYS APDL (*ANSYS® Academic Research, Release 17.0*), so that any topology pixel is subdivided into layers of four identical elements. As mentioned in Section 2, just half of unit-cell's thickness is modelled to decouple asymmetric modes, and for reduced computational cost of optimization, this half thickness is modelled by a single layer of elements for estimation and maximization of the bandgap of low order modal branches (Hedayatrasa et al., 2016b). Selected optimized results are then analyzed by 10 layers of linear solid elements with in-plane mesh resolution of $72 \times 72$.

A set of five different topologies are taken from the obtained Pareto fronts as shown in Figure 3(a) & (b); TS1 to TS5 for maximized stiffness and TM1 to TM5 for minimized mass. Unit-cell width of $a = 10$ mm (and so thickness of $h = 5$ mm) is assumed and relevant bandgap frequency range of selected topologies is shown in Figure 3(c) confined between its lower and upper limits. As expected, the bandgap frequency level of LRAB topologies $1.5 < f < 4.5$ kHz is quite low compared with assumed unit-cell width of 10 mm. The relative compliance of the topologies is also depicted and compared in Figure 3(d).

For the topologies with maximized stiffness, the topology TS1 is the Pareto extreme topology with maximum RBW which has the lowest stiffness. Selected intermediate topologies TS2 and TS3 have the same locally resonant core inclusion and cavity but different matrix designs. The RBW of topology TS2 with bi-material matrix is more than that of heavier topology TS3 with a fully copper matrix as its bandgap is widened on the upper side (Figure 3(c)) from 3.45 to 3.60 kHz, while their relative compliance is slightly different (Figure 3(d)).

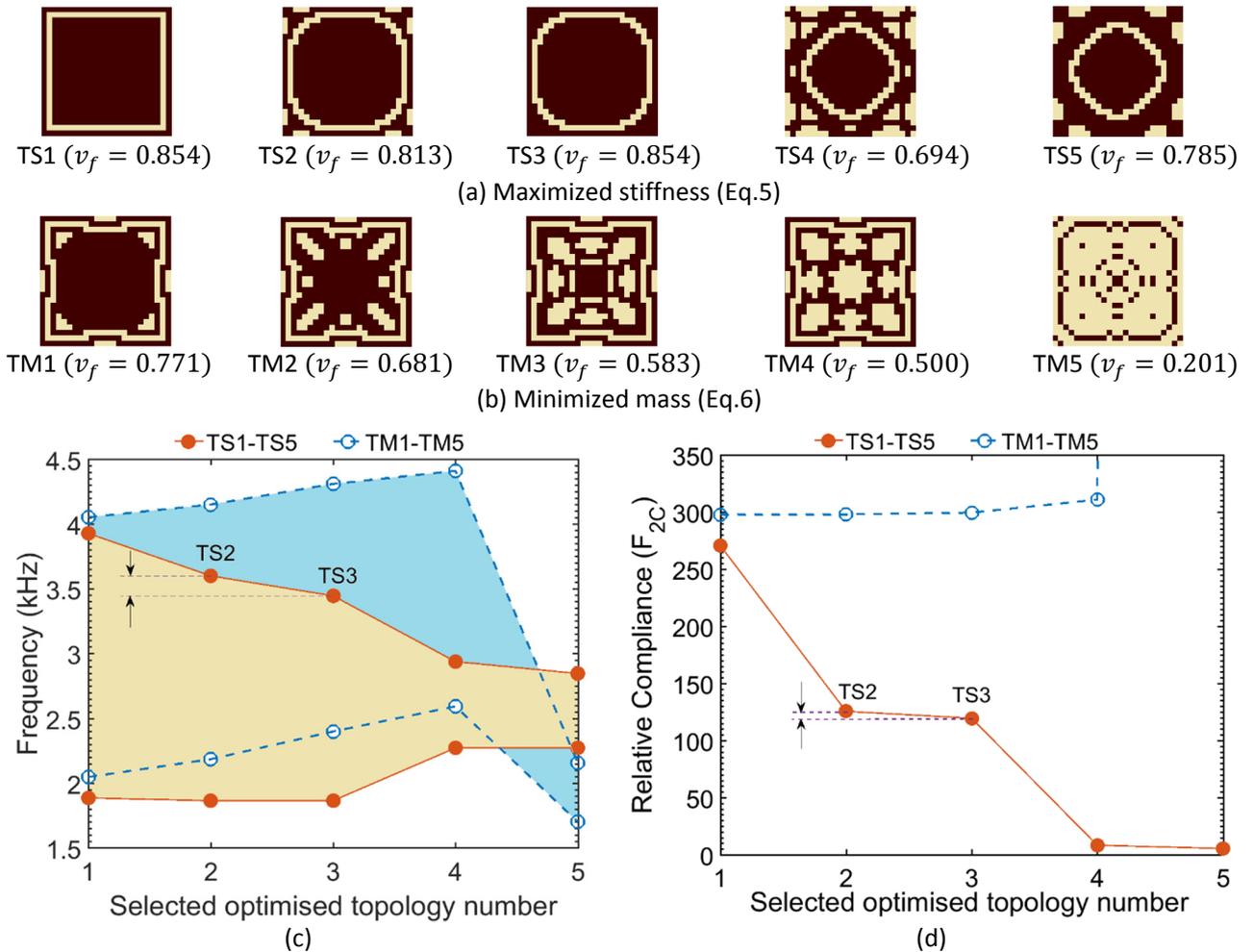

Figure 3 Selected optimized copper(dark)-rubber(light) LRAB topologies for (a) maximized stiffness and (b) minimized filling fraction of copper phase, (c) bandgap frequency range of selected topologies for unit-cell width of $a = 10$ mm (i.e. $h = 5$ mm), (d) relative compliance and filling fraction of selected topologies

Similarly, the bandgap of topology TS4 is widened on the upper side compared with that of TS5 through the same resonant and intermediate coating domain. Consequently, increased RBW and reduced weight can be achieved through minimal degradation of compliance by appropriate design optimization of the matrix (i.e. solid frame encompassing resonating core).

Regarding the optimization results for minimized mass, topologies TM1 and TM5 (Figure 3(b)) are the two extreme topologies of relevant Pareto front with maximum RBW (maximum mass) and minimum mass (minimum RBW), respectively. Extreme topology TM1 and selected intermediate topologies TM2 to TM4 have the same matrix design (and so slightly different compliances as shown in Figure 3(d)) but different locally resonant cores. Unlike the other set of topologies (TS1 to TS5), the resonant core in TM1 to TM4 is not fully copper and the mass minimization has inspired a bi-material design. Reduced mass of resonant core shows no noticeable change in the bandgap width of TM1 to TM4 but, as expected, increases the bandgap frequency range due to increasing resonance frequency. The other extreme topology TM5 with minimum mass shows a totally different topology mode with soft rubber matrix (and so very low relative compliance excluded from Figure 3(d)) which provides narrow and low-frequency bandgap.

The optimized topologies are comprised of separate domains which may be replaced by different materials to tune the bandgap. The isolated segments inside the resonant core and inside stiff matrix may also be removed to further reduce the weight of unit-cell and facilitate its fabrication. To study the matter, the two topologies TS4 ($v_f = 0.694$) and TM2 ($v_f = 0.681$) with almost the same filling fraction are further analyzed by considering three different material compositions as given in Table 1.

Table 1 Different unit-cell material compositions used for modal band analysis of topologies TS4 and TL2

| Composition | Stiff matrix | Locally resonant core | Compliant intermediate coating | Isolated inclusions inside matrix/core |
|---|---|---|---|---|
| C1 (Optimized) | Copper | Copper | Rubber | Rubber |
| C2 | Aluminum | Copper | Rubber | Rubber |
| C3 | Aluminum | Copper | Rubber | Void |

The three compositions are depicted in Figure 4(a) for a finite lattice of size $2 \times 2$ and relevant modal band structures are shown on the lower side. The composition C1 is the original bi-material one assumed for topology optimization. In composition C2 the copper matrix is replaced with aluminum ($E_{Al} = 70 \times 10^9$ Pa, $v_{Al} = 0.34$ and $\rho_{Al} = 2700$ kg/m$^3$), and subsequently in composition C3 the isolated segments of both matrix and resonant core are also removed. With original bi-material composition C1, the stiffer topology TS4 has considerably narrower bandgap compared with the topology TM2. In composition C2, replacing the copper matrix with aluminum pushes up the upper limit of bandgap in both topologies and increases RBW. Finally removing the isolated segments (islands) of the matrix and resonant core further widens the bandgaps on the upper side and increases RBW.

The first modal branch is fundamentally governed by the locally resonant features and insignificantly affected by the applied alterations of material composition. The second modal branch is but highly dependent on the matrix domain and therefore the topology TS4, with a larger fraction of replaced copper matrix and its removed rubber segments, is more significantly affected by the introduced alterations. Although the mass of resonant core in TM2 is reduced when shifting from composition C2 to C3, its first modal branch is not noticeably affected and the bandgap is even slightly widened on the upper side due to minor porosities introduced in its matrix domain. The results agree with the trampoline design proposed by Bilal and Hussein (2013) in which bandgap efficiency of a plate with stub resonators was enhanced through periodic circular perforations.

The topology resolutions may be refined by an appropriate approach (Hedayatrasa et al., 2016a) or a level set-based topology optimization may be performed to determine optimal topology shape of achieved domains with multi-material composition. The thickness of coating layer should be set to the desired size as maximization of RBW naturally tends to enlarge the core and narrow this coating layer.

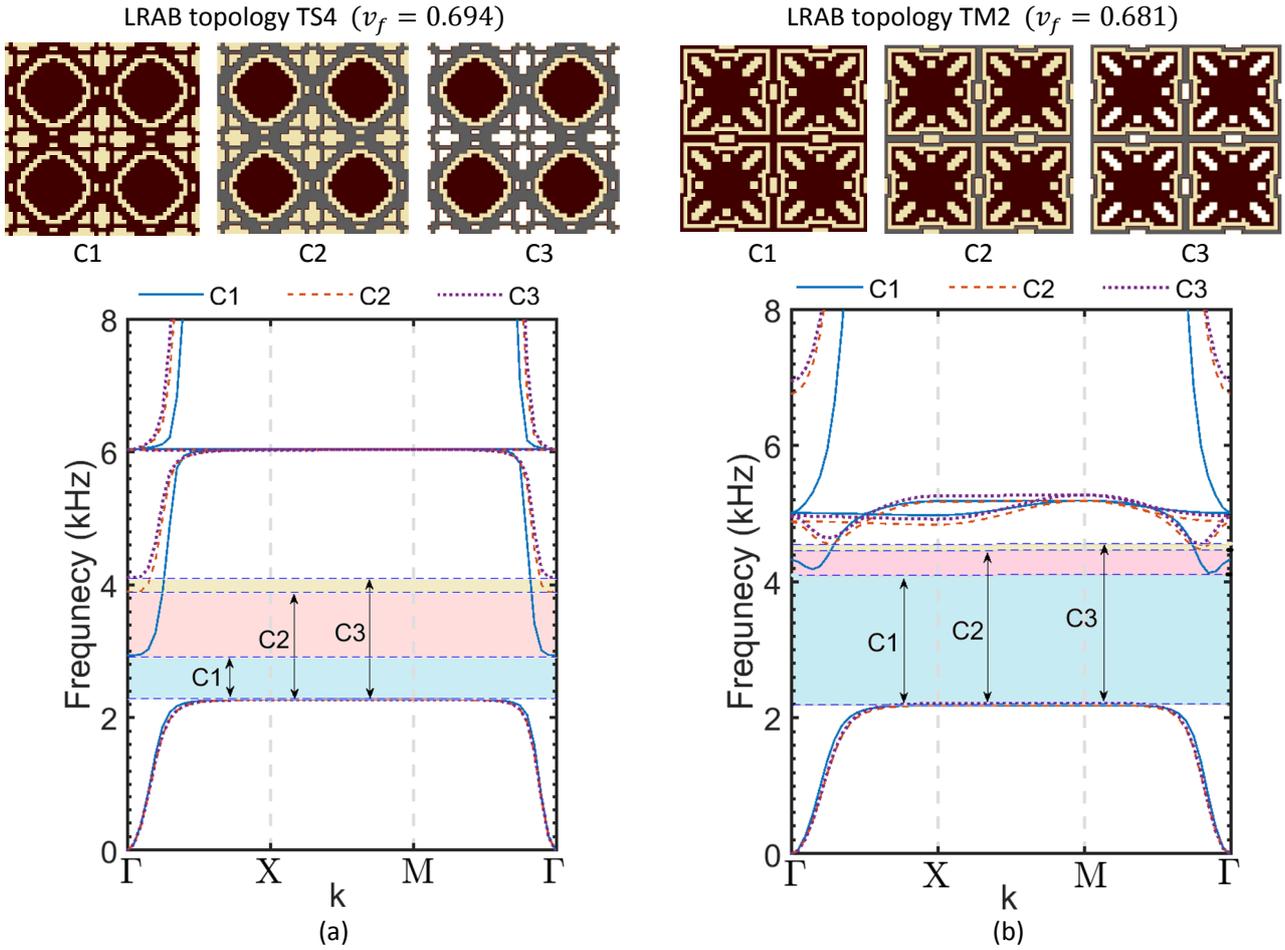

Figure 4 Finite 2 × 2 lattice of LRAB topology, alteration of material composition (Table 1) and relevant modal band structures for unit-cell width of $= 10$ mm , (a) topology TS4 and (b) topology TM2

## 4. CONCLUSIONS

Topology optimization study was performed for maximized locally resonant bandgap efficiency via a bi-material constitution. Maximized effective stiffness and minimized mass were also individually introduced as second objective and selected topologies of obtained Pareto fronts were evaluated. Soft rubber and stiff-dense copper were used as the constitutive materials during optimization and their distinct contrast inspired appearance of core-shell like locally resonant bandgaps for maximized relative bandgap efficiency. In contrary to the conventional topology of core-shell design, (i) maximizing bandgap through maximized stiffness introduced non-homogeneous matrix domain and maintained a homogeneous dense core, and (ii) maximizing bandgap through minimized mass dominantly inspired non-homogeneous resonant core. Non-homogeneity of the matrix widens the bandgap on the upper limit and reduces the overall mass while minimal degradation of effective structural stiffness is ensured through implemented optimization. Moreover, for specified filling fraction of stiff copper material in the resonant core, non-homogeneous core design offers an enlarged peripheral boundary inside the matrix's cavity leading to higher bandgap efficiency. Multi-material designs were also evaluated by altering the composition of selected optimized topologies. Replacing copper matrix with aluminum increased the upper limit of bandgap, and porosities introduced by removing isolated rubber segments of topology further widened the bandgap efficiency.

The results surprisingly suggest that, for any given material composition, the locally resonant bandgap efficiency can be further enhanced by microstructural design optimization of the matrix and the resonant core while maximizing the effective stiffness and/or minimizing the overall mass.